\documentclass{article}
\usepackage[preprint]{spconfa4}
\usepackage{amsmath,graphicx,cite}

\usepackage{xcolor,array}
\usepackage{tikz,booktabs}
\usepackage{nicefrac}

\newcommand{\VEC}[1]{\mathbf{#1}}          

\newcommand{\putindex}[3]{\vtop{\hbox{\hspace{#3} $#1$}
            \hbox{\raise 6mm \hbox{$\scriptscriptstyle #2$}}}}

\newcommand{\gradx}[0]{\vtop{\hbox{\rm grad}
            \hbox{\raise 2.5mm \hbox{\rm \hspace{2mm} \footnotesize x}}}}

\newcommand{\grady}[0]{\vtop{\hbox{\rm grad}
            \hbox{\raise 2.5mm \hbox{\rm \hspace{2mm} \footnotesize y}}}}

\newcommand{\grad}[1]{\vtop{\hbox{\rm grad}
            \hbox{\raise 2.5mm \hbox{#1}}}}

\newcommand{\btb}{     \begin{tabbing}             }
\newcommand{\bte}{     \end{tabbing}               }

\usetikzlibrary{dsp}
\usetikzlibrary{chains}
\usetikzlibrary{shapes.geometric}

\graphicspath{ {./placeholder/} }
\usetikzlibrary{external}

\title{BANDWIDTH-SCALABLE FuLLY MASK-BASED DEEP FCRN \\ ACOUSTIC ECHO CANCELLATION AND POSTFILTERING}
\name{\parbox{\linewidth}{\centering Ernst Seidel$^{\ast}$, Rasmus Kongsgaard Olsson$^{\circ}$, Karim Haddad$^{\circ}$, Zhengyang Li$^{\ast}$, \mbox{Pejman Mowlaee$^{\circ}$, Tim Fingscheidt$^{\ast}$}}}
\address{$^{\ast}$Institute for Communications Technology,
	Technische Universität Braunschweig\\
	Schleinitzstraße 22,
	38106 Braunschweig, Germany\\ $^{\circ}$GN Audio A/S,
	Lautrupbjerg 7,
	2750 Ballerup, Denmark}

\begin{document}
\ninept
\maketitle
\begin{abstract}
Although today’s speech communication systems support various bandwidths from narrowband to super-wideband and beyond, state-of-the art DNN methods for acoustic echo cancellation (AEC) are lacking modularity and bandwidth scalability. Our proposed DNN model builds upon a fully convolutional recurrent network (FCRN) and introduces scalability over various bandwidths up to a fullband (FB) system (48 kHz sampling rate). This modular approach allows joint wideband (WB) pre-training of mask-based AEC and postfilter stages with dedicated losses, followed by separate \mbox{trainings} of them on FB data. A third lightweight blind bandwidth extension stage is separately trained on FB data, flexibly allowing to extend the WB postfilter output towards higher bandwidths until reaching FB. Thereby, 
higher frequency noise and echo are reliably suppressed. On the ICASSP 2022 Acoustic Echo Cancellation Challenge blind test set we report a competitive performance, showing robustness even under highly delayed echo and dynamic echo path changes.

\end{abstract}
\begin{keywords}
acoustic  echo  cancellation, bandwidth extension, echo suppression, convolutional LSTM
\end{keywords}

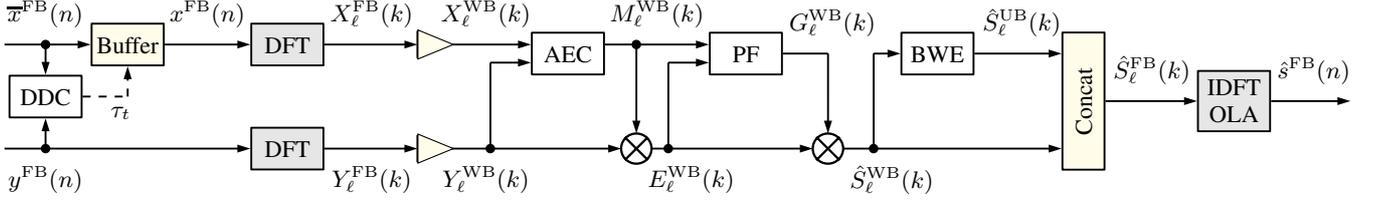
\begin{figure*}[t!]
	\centering
  \tikzsetnextfilename{figures/model_overview}%
  \begin{tikzpicture}[scale = 0.6]
	
	\draw   (0.0,0.0)   coordinate    (startX)    {}
	+       (0.0,-2.3)  coordinate    (startY)    {}
	++      (0.9,0.0)   node    (xDel)      [dspnodefull] {}
	+       (0.0,0.0)   node    (t_x)       [label=above:${\overline{x}}^\mathrm{FB}(n)$] {}
	+       (0.0,-1.15) node    (delay)     [dspfilter, minimum width=3em, minimum height=1.75em] {DDC}
	+       (1.65,-1.0) node   (t_d)       [label=below:${\tau}_t$] {}
	+       (0.0,-2.3)  node    (yDel)      [dspnodefull] {}
	+       (0.0,-2.3)   node    (t_y)       [label=below:${y}^\mathrm{FB}(n)$] {}
	++      (1.8,0.0)   node    (ring)      [dspfilter, fill=yellow!10, minimum width=3em, minimum height=1.75em] {Buffer}
	+       (1.75,0.0)   node    (t_xd)       [label=above:${x}^\mathrm{FB}(n)$] {}
	+       (0.0,-1.15) coordinate   (d_tau)     {}
	++      (3.5,0.0)   node    (xDFT)      [dspfilter, fill=black!10, minimum width=3em, minimum height=1.75em] {DFT}
	+       (0.0,-2.3)  node    (yDFT)      [dspfilter, fill=black!10, minimum width=3em, minimum height=1.75em] {DFT}
	++      (1.85,0.0)   node    (t_XFB)     [label=above:${X}^\mathrm{FB}_\ell(k)$] {}
	+       (0.0,-2.3)   node    (t_XFB)     [label=below:${Y}^\mathrm{FB}_\ell(k)$] {}
	++      (1.0,0.0)   coordinate    (xTri)      {}
	+       (0.7,0.0)   coordinate    (xTriE)     {}
	+       (0.0,-2.3)  coordinate   (yTri)      {}
	+       (0.7,-2.3)  coordinate   (yTriE)     {}
	++      (1.5,0.0)   node    (t_XWB)     [label=above:${X}^\mathrm{WB}_\ell(k)$] {}
	+       (0.0,-2.3)  node    (t_XWB)     [label=below:${Y}^\mathrm{WB}_\ell(k)$] {}
	++      (1.8,-0.2)  node    (AEC)       [dspfilter, minimum width=3em, minimum height=1.75em] {AEC}
	+       (-1.7,-2.1) node    (y_WB)      [dspnodefull] {}
	++      (1.5,0.2)   node    (m_WB)      [dspnodefull] {}
	+       (0.45,0.0)   node    (t_M)       [label=above:${M}^\mathrm{WB}_\ell(k)$] {}
	+       (0.0,-2.3)  node    (mult1)     [dspmixer, fill=white] {}
	++      (2.4,-0.2)   node    (PF)       [dspfilter, minimum width=3em, minimum height=1.75em] {PF}
	+       (-1.7,-2.1) node    (e_WB)      [dspnodefull] {}
	+       (-1.2,-2.1)   node    (t_e)       [label=below:${E}^\mathrm{WB}_\ell(k)$] {}
	++      (1.8,0.0)   coordinate    (g_WB)      {}
	+       (0.1,0.0)   node    (t_G)       [label=above:${G}^\mathrm{WB}_\ell(k)$] {}
	+       (0.0,-2.1)  node    (mult2)     [dspmixer, fill=white] {}
	++      (1,0.0)   coordinate    (BWE_L)     {}
	+       (0.0,-2.1)  node    (s_WB)      [dspnodefull] {}
	+       (0.4,-2.1)   node    (t_s)       [label=below:$\hat{{S}}^\mathrm{WB}_\ell(k)$] {}
	++      (1.4,0.0)   node    (BWE)       [dspfilter, minimum width=3em, minimum height=1.75em] {BWE}
	++      (1.25,0.0)   coordinate    (BWE_R)     {}
	+       (0.6,0.0)   node    (t_BWE)       [label=above:$\hat{{S}}^\mathrm{UB}_\ell(k)$] {}
	++      (1.95,-1.05)  node    (conc)      [dspfilter, rotate=90, fill=yellow!10, minimum width=5.75em, minimum height=1.75em] {Concat}
	+       (1.5,0.0)   node    (t_BWE)       [label=above:$\hat{{S}}^\mathrm{FB}_\ell(k)$] {}
	++      (3.3,0.0)   node    (IDFT)      [dspfilter, fill=black!10, minimum width=3em,  text height=2.1em] {IDFT \\ OLA}
	+       (1.75,0.0)   node    (t_BWE)       [label=above:$\hat{{s}}^\mathrm{FB}(n)$] {}
	++      (2.55,0.0)   coordinate    (end)       {};
	
	\draw ([yshift=+10.5mm] conc.north) coordinate (concu);
	\draw ([yshift=-10.5mm] conc.north) coordinate (concd);
	
	\draw ([yshift=+19mm] y_WB)     coordinate (y_WB_u);
	\draw ([yshift=+2mm] AEC.west)  coordinate (AECu);
	\draw ([yshift=+2mm] AEC.east)  coordinate (AECo);
	\draw ([yshift=-2mm] AEC.west)  coordinate (AECd);
	
	\draw ([yshift=+19mm] e_WB)     coordinate (e_WB_u);
	\draw ([yshift=-2mm] PF.west)   coordinate (PFd);
	\draw ([yshift=+2mm] PF.west)   coordinate (PFu);
	
	\draw ([yshift=+3mm] xTri) coordinate (xTriU);
	\draw ([yshift=-3mm] xTri) coordinate (xTriD);
	\draw ([yshift=+3mm] yTri) coordinate (yTriU);
	\draw ([yshift=-3mm] yTri) coordinate (yTriD);
	
	\begin{scope}[start chain]
	
	\chainin (xTriU);
	\chainin (xTriD)    [join=by dspline];
	\chainin (xTriE)    [join=by dspline];
	\chainin (xTriU)    [join=by dspline];
	\chainin (yTriU);
	\chainin (yTriD)    [join=by dspline];
	\chainin (yTriE)    [join=by dspline];
	\chainin (yTriU)    [join=by dspline];
	
	\chainin (startX);
	\chainin (ring) 	[join=by dspconn];
	\chainin (xDFT) 	[join=by dspconn];
    \chainin (xTri) 	[join=by dspconn];
    
    \begin{scope}[dashed]
        \chainin (delay);
        \chainin (d_tau)    [join=by dspline];
        \chainin (ring)    [join=by dspconn];
    \end{scope}
    
	\chainin (xTriE);
	\chainin (AECu) 	[join=by dspconn];
    \chainin (AECo);
	\chainin (PFu) 	    [join=by dspconn];
	\chainin (PF);
	\chainin (g_WB) 	[join=by dspline];
	\chainin (mult2) 	[join=by dspconn];
	
	\chainin (startY);
	\chainin (yDFT) 	[join=by dspconn];
    \chainin (yTri) 	[join=by dspconn];
    
	\chainin (yTriE);
	\chainin (mult1) 	[join=by dspconn];
	\chainin (mult2) 	[join=by dspconn];
	\chainin (concd) 	[join=by dspconn];
	\chainin (conc);
	\chainin (IDFT) 	[join=by dspconn];
	\chainin (end) 	    [join=by dspconn];
	
	\chainin (yDel);
	\chainin (delay) 	[join=by dspconn];
	\chainin (xDel);
	\chainin (delay) 	[join=by dspconn];
	
	\chainin (y_WB);
	\chainin (y_WB_u) 	[join=by dspline];
	\chainin (AECd) 	[join=by dspconn];
	
	\chainin (m_WB);
	\chainin (mult1) 	[join=by dspconn];
	
	\chainin (e_WB);
	\chainin (e_WB_u) 	[join=by dspline];
	\chainin (PFd) 	    [join=by dspconn];
	
	\chainin (s_WB);
	\chainin (BWE_L) 	[join=by dspline];
	\chainin (BWE) 	    [join=by dspconn];
	\chainin (concu) 	[join=by dspconn];

    \end{scope}
    
    \draw (xTriE) -- (xTri) node[draw,isosceles triangle, fill=yellow!10,anchor=west]{};
    \draw (yTriE) -- (yTri) node[draw,isosceles triangle, fill=yellow!10,anchor=west]{};
	
\end{tikzpicture}%

    \vspace{-0.8cm}
	\caption{{Proposed model}, comprising DDC, AEC, PF, and BWE. The ringbuffer is labelled "Buffer", the concatenation of PF and BWE output is labelled as "Concat", and the triangular operators mark our bandwidth-limiting functions zeroing out upper band frequencies ($8...24$ kHz).}
	\label{fig:overview}
	\vspace{-5.0mm}
\end{figure*}

\section{Introduction}
\label{sec:intro}

Speech communications today is offered in various bandwidths ranging from narrowband to fullband (sampling frequency $f_s = 48$ kHz). This can be prominently seen from the Enhanced Voice Services (EVS) speech codec \cite{3GPP_26090}. In consequence, also acoustic echo cancellation must support these bandwidths, posing the challenge of a bandwidth-scalable design. In the EVS codec, as in the AMR codec \cite{3GPP_26190}, interestingly, for speech transmission there is a core operating at $f_s=12.8$ kHz (cut-off frequency $6.4$ kHz), while higher frequencies are estimated by (blind) bandwidth extension techniques. Our work takes on this approach, providing a modular and bandwidth-scalable solution to deep acoustic echo cancellation (AEC) and postfiltering (PF) employing blind bandwidth extension beyond 8 kHz.

Deep AEC studies can be divided into two groups: {\rm single-stage methods} \cite{wang_NN_AEC_19,westhausen20_interspeech,zhang21ia_interspeech} naturally prohibiting a separate loss for AEC and residual echo suppression PF, and {\rm multi-stage methods} \cite{carbajal_RES_ICASSP,Valin_AEC,pfeifenberger21_interspeech,peng21f_interspeech,seidel21_interspeech}, which consist of an AEC stage to remove major parts of the echo followed by a DNN PF to suppress residual echo and noise. Multi-stage methods can be further divided into two subgroups:\!\! {Either \mbox{(i) both} AEC and PF are DNNs (\textrm{DNN AEC/PF}) \cite{seidel21_interspeech}, or (ii) a linear AEC \cite{SpeexDSP,Enzner2006} is followed by a DNN PF (\textrm{LAEC/DNN PF}) \cite{carbajal_RES_ICASSP,Valin_AEC,pfeifenberger21_interspeech,peng21f_interspeech}.}

Promising results have been reported in wideband AEC challenges \cite{Sridhar2021,cutler2021} from hybrid solutions. These works outperformed the respective single-stage mask-based baseline method in terms of AEC-MOS \cite{purin2021aecmos} and subjective evaluation \cite{cutler2021crowdsourcing}. It was confirmed in \cite{Franzen2021} that the quality of the nearend speech component may degrade when using a single-stage DNN AEC/PF, concluding that it is a better choice to have a separate AEC DNN. Accordingly, in \cite{seidel21_interspeech} a two-stage DNN AEC/PF approach was proposed, where the first stage directly estimates an echo that is then {\it subtracted} from the microphone signal (motivated from classical hands-free systems), followed by a second-stage mask-based DNN PF. Opposed to a single-stage DNN, the two-stage concept reduces the task complexity per stage. Still, the employed first-stage direct echo estimation is at a disadvantage compared to mask-based approaches since it requires reconstruction of the echo's absolute spectral amplitudes.

In this paper, we propose a bandwidth-scalable design for speech enhancement applications which relies on three modular DNN stages: AEC, PF, and bandwidth extension. The novelty of this approach is threefold: 
(i) \textit{Fully mask-based speech enhancement}: The first two stages are built upon our previous Y$^2$-Net model \cite{seidel21_interspeech}, with the DNN AEC now also predicting a complex mask.
(ii) \textit{Bandwidth scalability}: Our core AEC and PF modules can operate on WB speech, but with a lightweight bandwidth extension also on higher bandwidths up to FB. 
(iii) \textit{Multiple separate processing stages (modularity)}: 
The AEC and PF modules are pre-trained on WB but later jointly trained on FB data, while the bandwidth extension module is independently trained on FB data. 
Please note that to the best of our knowledge, we are the first to propose a  fully mask-based modular and bandwidth-scalable approach and this is the first FB AEC proposal among the multi-stage DNN AEC/PF methods.

The remainder of this paper is structured as follows: In Section 2, a system overview including the framework and its individual stages is given. The training and different experimental variants including choices of each stage are described in Section 3. In Section 4, the experimental results of our proposed method are discussed. Section 5 provides conclusions.
\section{System Overview and Proposed Method}
\label{sec:method}

\subsection{Dynamic Delay Compensation (DDC)}
\label{subsec:delay}
As the delay between the far-end (FE) and near-end (NE) signal is generally unknown, in this work we propose employing dynamic delay compensation (DDC) via applying the GCC-PHAT algorithm \cite{KnappCarter1976} to find the time delay between the two signals and to then delay the far-end signal by the estimated time delay. Previous studies \cite{zhang21ia_interspeech,pfeifenberger21_interspeech,peng21f_interspeech} have reported that a time alignment pre-processing stage helps to improve the AEC performance and to increase the robustness of the subsequent processing stages against delayed echo signals.

Given that a 48 kHz sampled (fullband) microphone signal ${y}^\mathrm{FB}(n)={s}^\mathrm{FB}(n)+{d}^\mathrm{FB}(n)+{n}^\mathrm{FB}(n)$ consists of the components NE speech ${s}^\mathrm{FB}(n)$, echo ${d}^\mathrm{FB}(n)$, and noise ${n}^\mathrm{FB}(n)$, the goal of the DDC is to find the correlation between ${d}^\mathrm{FB}(n)$ and the FE \mbox{(= loudspeaker)} signal $\overline{{x}}^\mathrm{FB}(n)$. The algorithm uses frames of  ${y}^\mathrm{FB}(n)$ and $\overline{{x}}^\mathrm{FB}(n)$ with $n \in \mathcal{N}_t$ and $|\mathcal{N}_t| = 50,880$ ($\widehat{=}\, 1.06$\,s length) and a 25\% frame shift. This corresponds to 40 full frames of the subsequent processing stages. The {\it instantaneous} delay $\overline{\tau}_t$ is estimated for each DDC frame with index $t$ by applying a $\overline{K} = 50,880$-point DFT and computing $\left( \text{DFT bin } \overline{k} \in \overline{\mathcal{K}} \right)$

\vskip-12pt

\begin{align}
	\Phi_t \!\left(\overline{k}\right) & = \alpha \Phi_{t\!-\!1}\!\left(\overline{k}\right) + (1\!-\!\alpha) Y^\mathrm{FB}_t \!\left(\overline{k}\right)\!\cdot\! \left(\overline{X}_t^\mathrm{FB} \!\left(\overline{k}\right)\right)^{\!*} \!\!, \ \ \overline{k}\in \mathcal{\overline{\overline{K}}}\label{eq:phi} \\
	\overline{\tau}_t & = \arg\max_\tau \phi_t(\tau)\quad \textrm{with}\quad \phi_t(\tau) = 
	{\rm IDFT}\frac{\Phi_t\!\left(\overline{k}\right)}{|\Phi_t\!\left(\overline{k}\right)|}, \label{eq:tau}
\end{align}
where $Y^\mathrm{FB}_t \!\left(\overline{k}\right)$ is the microphone signal DFT and $\left(\overline{X}_t^\mathrm{FB} \!\left(\overline{k}\right)\right)^*$ is the complex conjugated far-end signal DFT. Note that (\ref{eq:phi}) is computed for $\overline{k}\in \mathcal{\overline{\overline{K}}}$, where $ \mathcal{\overline{\overline{K}}} \subset \mathcal{\overline{K}}$
limits the observed bandwidth to a range of $200$ Hz to \mbox{$8$ kHz} for fast and robust estimation. 
The smoothing factor $\alpha$ is set to $0.7$ to achieve a good balance between accuracy and responsiveness towards delay changes. 
{Delay changes are only evaluated after the following DDC frame shift, leaving $0.265$\,s to compute $\overline{\tau}_t$ in parallel to other processing stages and therefore without adding additional latency to the system. Since the subsequent processing stages are already robust towards delays, a relaxed configuration is chosen for computing the {\it actual active delay} estimate $\tau_t$. If a stable $\overline{\tau}_t$ is detected for $2$ consecutive frames, $\tau_t$ will be updated to the current $\overline{\tau}_t$ reduced by $200$\,ms (down to a minimum of $0$\,ms). The $200$\,ms buffer minimizes the risk of non-causal delay paths caused by over-estimations or not yet identified delay changes. As shown in Figure \ref{fig:overview}, the estimated delay is compensated on the reference path using a ringbuffer that produces  \mbox{$x^\mathrm{FB}(n) =  \overline{x}^\mathrm{FB}(n - \tau_t$)}. With parallel computation of $\overline{\tau}_t$ and the stability requirement, changes in the delay can be detected with an algorithmic reaction time of $0.53$\,s.}

\subsection{Deep AEC and Postfilter}
\label{subsec:FCRN}

After time-alignment by the DDC, both the FB microphone signal and the delay-compensated FB reference signal are passed through a 
1st order high-pass filter with 50\,Hz cut-off frequency to eliminate time-varying biases \cite{Halimeh2020}. The resulting signals ${y}^\mathrm{FB}(n)$ and ${x}^\mathrm{FB}(n)$ are subject to square root Hann windowing of frame length \mbox{$L = 1,\!272$} with a frame shift of 50\%. The frames are zero-padded for transformation into the frequency domain using a \mbox{$K=1,\! 536$-}point DFT. From the frequency domain representations $X^\mathrm{FB}_\ell(k)$ and $Y^\mathrm{FB}_\ell(k)$ with frame index $\ell$ and frequency bin index \mbox{$k \in \mathcal{K}=\{0,1,...,K$\!$-$\!$1\}$}, only the frequency bins \mbox{$k \in \mathcal{K}' \subset \mathcal{K}$} up to $8$ kHz are passed on to the following processing stages (effectively WB speech), while the upper band (UB) bins are simply set to zero (depicted in Figure \ref{fig:overview} by a triangular operator in each signal path). This method greatly reduces the complexity required by our processing stages. Higher frequencies are later restored by the BWE stage as these bands are more likely to be dominated by noise rather than NE speech components, which makes direct denoising harder. 
This system design also does not rely on a fixed sampling rate of $48$\,kHz, or the presence of actual signal frequency components above $8$\,kHz, and can be used with lower sampling rates if desired as long as the  frame length $L$ covers the same time span of $T = 26.5\,\mathrm{ms}$ and DFT lengths and discarded frequency bins are adjusted accordingly. Different sampling rates only require separate weight sets.

The proposed AEC and PF stages build upon the so-called \mbox{Y$^2$-Net} architecture \cite{seidel21_interspeech}. The AEC stage is trained to perform only echo cancellation, while the PF stage suppresses residual echo as well as noise. All network configuration parameters such as strides, kernel sizes and kernel numbers remain unchanged. Both stages maintain the same structure of a convolutional LSTM placed in the bottleneck of an convolutional encoder-decoder structure, thereby being a fully convolutional recurrent network (FCRN) after Strake et al.~\cite{Strake2020b}. There are separate encoder paths in the AEC stage (referred to as late fusion in the Y$^2$-Net). A structural novelty of this work consists in the AEC stage output, which is now an {\it estimated complex mask} 
$M^\mathrm{WB}_\ell(k)$ with {\it compressed amplitude} before applying it to $Y^\mathrm{WB}_\ell(k)$, according to

\vskip-12pt

\begin{equation}
    {E}_\ell^\mathrm{WB}(k) = Y_\ell^\mathrm{WB}(k) \cdot \tanh(|M_\ell^\mathrm{WB}(k)|) \cdot\frac{M_\ell^\mathrm{WB}(k)}{|M_\ell^\mathrm{WB}(k)|}. \label{eq:mask}
\end{equation}
For the PF stage, a further structural novelty is that besides the usual echo-reduced input ${E}_\ell^\mathrm{WB}(k)$, also the AEC mask $M^\mathrm{WB}_\ell(k)$ will be used as secondary input. The estimated complex PF mask $G^\mathrm{WB}_\ell(k)$ is applied to ${E}_\ell^\mathrm{WB}(k)$ analog to (\ref{eq:mask}).

Since the proposed model is designed in a modular fashion, the BWE stage can be omitted if only WB signal components are required as output. In this case, the PF output $\hat{S}_\ell^\mathrm{WB}(k)$ is synthesized into a time-domain signal $\hat{\VEC{s}}^\mathrm{WB}$ using a frame-wise IDFT, {square-root Hann synthesis window} and overlap-add (OLA) with the same parameters as used for analysis. This results in a total algorithmic delay of $T_1 = T + T_s$. With our chosen frame length $T=26.5\,\mathrm{ms}$ and frame shift $T_s=13.25\,\mathrm{ms}$, we achieve $T_1=39.75\,\mathrm{ms}$ with an additional negligible delay caused by the initial high-pass filter.

\subsection{Bandwidth Extension}
\label{subsec:BWE}
To restore FB signals, one may add a third processing stage, a BWE method. In our proposal, this is a lightweight algorithm that directly uses $\hat{S}_\ell^\mathrm{WB}(k)$ as input. The signal is separated into amplitude $\big|\hat{S}_\ell^\mathrm{WB}(k)\big|$ and phase $\phi(\hat{S}^\mathrm{WB}_\ell(k))$. A simple DNN that is similar to the works of Abel et al.~\cite{Abel2018} is employed to estimate the UB amplitudes $\big|\hat{S}_\ell^\mathrm{UB}(k)\big|$. This DNN takes $\ln\big|\hat{S}_\ell^\mathrm{WB}(k)\big|$ as input to three fully connected layers, featuring 256 nodes and ReLU activation each, followed by an output layer of 512 nodes with linear activation, and an exponential function to return from the logarithmic scale. To prevent overly strong estimations, especially on residual echo, we introduced an attenuation factor $\gamma_\ell = \min ( 1,\, \theta \sqrt{P_\ell^\mathrm{WB}/P_\ell^\mathrm{UB}})$, with $P_\ell$ being the average spectral bin power of the upper band (UB) or the lower band (WB), respectively, and $\theta = 0.1$ chosen from informal listening tests.
The upper band phases $\phi(\hat{S}^\mathrm{UB}_\ell(k))$ are created by simply concatenating copies of  $\phi(\hat{S}^\mathrm{WB}_\ell(k))$
twice, thereby extending from $8$ kHz to $24$ kHz. The complex upper band frequency bins $\hat{S}_\ell^\mathrm{UB}(k) = \gamma_\ell \big|\hat{S}_\ell^\mathrm{UB}(k)\big|\cdot\phi(\hat{S}^\mathrm{UB}_\ell(k))$ are then concatenated with $\hat{S}_\ell^\mathrm{WB}(k)$ to form the bandwidth-extended FB signal $\hat{S}_\ell^\mathrm{FB}(k)$, which can then be transformed back into the time domain in analogy to the description in Section \ref{subsec:FCRN}. This implementation of BWE introduces no additional algorithmic delay.

The entire proposed system is of similar complexity and real-time capabilities as the \mbox{Y$^2$-Net}, as verified on an {\tt Intel i7-10510U} quad-core CPU at $1.8$\,GHz. The DDC algorithm requires only few resources which can be spread out over a full delay update period, and the BWE-DNN only adds 14k more parameters to the combined 7.5M of AEC and PF.
\section{Experimental Setup}
\label{sec:setup}

\subsection{Pre-Training: Wideband AEC and PF}
\label{subsec:pretraining}

Since the FB AEC and PF processing stages effectively only operate on signals in the frequency range $0...8$ kHz (WB), we perform a pre-training of weights on $16$\! kHz sampled data using window length $L'=L/3=424$ and DFT size $K'=K/3=512$. 
As with the original WB \mbox{Y$^2$-Net} approach \cite{seidel21_interspeech}, these pre-training steps were performed on the Interspeech 2021 AEC Challenge synthetic dataset~\cite{cutler2021}, which consists of 10,000 WB audio samples of 10s length each. From these, 8,000 files are used for pre-training and 1,500 for validation between epochs. The pre-training, as well as all following training, are conducted using {\tt TensorFlow2}~\cite{Abadi2016}. With the novel double-masking approach and PF mask input, the employed frequency-domain MSE losses are defined as

\vskip-12pt

\begin{align}
{J}_{\ell}^\mathrm{AEC} & = \mathrm{MSE}({E}^\mathrm{WB, noisy}_\ell(k'),\,\, {S}^\mathrm{WB, noisy}_\ell(k') \label{eq:J_AEC}) \\
{J}_{\ell}^\mathrm{PF} & = \mathrm{MSE}(\hat{{S}}^\mathrm{WB}_\ell(k'),\,\, {S}^\mathrm{WB}_\ell(k')) \label{eq:J_PF}
\end{align}
using $\mathrm{MSE}(A_\ell(k'), B_\ell(k')) = \frac{1}{K'} \sum_{k' \in \mathcal{K}'} \big| {A}_\ell(k') - {B}_\ell(k')|^2$.
Here, $S_\ell^\mathrm{WB, noisy}(k')$ is a noisy target containing NE speech and noise components. The losses are then averaged across all sequence frames $\ell$ yielding $J^\mathrm{AEC}$ and $J^\mathrm{PF}$, followed by averaging over entries in the minibatch. In a first pre-training step we only train the AEC stage using $J^\mathrm{AEC}$. In a second pre-training step, both AEC and PF are jointly trained using the weighted loss term ($\alpha = 0.25$)

\vskip-3pt

\begin{equation}
	J  = \alpha J^\mathrm{AEC} + (1-\alpha) J^\mathrm{PF}. \label{eq:J}
\end{equation}
Both pre-training steps are conducted on WB data at $16$\,kHz sample rate, with a sequence length of $L^\mathrm{pre}=50$ frames and mini-batch size $B=16$. An Adam optimizer~\cite{adam_optimizer} is used with an initial learning rate of \mbox{$1\!\cdot\! 10^{-4}$}, which is decreased by a factor 0.5 after 4 epochs of stagnating validation losses, down to a minimum of $1\!\cdot\!10^{-5}$.

\subsection{Training: Wideband AEC and PF}
\label{subsec:trainWB}

The actual model training steps are now conducted on FB data using our proposed $48$\,kHz DFT-based analysis/synthesis scheme as shown in Fig.\,\ref{fig:overview}.$ $ Again, AEC and PF are trained in two separate steps, now, however, on the ICASSP 2022 AEC Challenge synthetic FB dataset~\cite{cutler2022AEC}, further referenced as $\mathcal{D}_\mathrm{syn}$, which also consists of 10,000 files of $10$\,s length. Compared to the pre-training dataset, $\mathcal{D}_\mathrm{syn}$ features more diversity, including speech-to-echo ratios (SERs) of $-25$\,dB $... 40$\,dB, noisy far-end references, varying bandwidth of microphone signal components and echo delays up to $300$\,ms. Of this dataset, the first 500 files are not used in training and form the preliminary synthetic test set $\mathcal{D}_\mathrm{syn}^\mathrm{test,MS}$ later used for BWE evaluation. The remaining files are split into 8,000 for training ($\mathcal{D}_\mathrm{syn}^\mathrm{train}$) and 1,500 for validation between epochs ($\mathcal{D}_\mathrm{syn}^\mathrm{val}$).

The training steps use a similar configuration as the pre-training steps. The first training step, tuning weights on the AEC stage, uses a sequence length of $L^\mathrm{train} = 200$, due to limited GPU memory requiring a small mini-batch size of $B=4$. This increased sequence length supports the AEC in learning time-context over longer periods. Furthermore, a time-domain logarithmic MSE loss is obtained for each entry in the mini-batch as {($\varepsilon$ small number)}

\vskip-12pt

\begin{align}
    {J}^{\mathrm{T-logMSE}} & = 10\cdot\log \left({\varepsilon} + 
    \mathrm{MSE}(\hat{{s}}^\mathrm{WB}(n'),\,\, {s}^\mathrm{WB}(n'))  \right), \label{eq:tlmse}
\end{align}
whereby MSE() sums over time instants $n'$ of the entire training sequence after synthesizing $\hat{{s}}^\mathrm{WB}(n)$ through IDFT of \mbox{$K$-length $\hat{S}_\ell(k)$} (UB frequency bins are set to zero), synthesis windowing, and overlap-add of all frames $\ell\in \{1,...,L^\mathrm{train}\}$.

The second training step again uses $L^\mathrm{train}=50$ and $B=16$, {\it but only adjusts weights within the PF stage}. Based on the findings by Braun et al.\ \cite{braun2021}, for this second training step we use magnitude-compressed (mC) and complex-compressed (cC) loss terms

\vskip-12pt

\begin{align}
    {J}_\ell^{\mathrm{mC}} & = 
    \mathrm{MSE}(|\hat{S}_\ell(k)|^c,\,\, |S_\ell(k)|^c) \label{eq:magcomp} \\
    {J}_\ell^{\mathrm{cC}} & = 
    \mathrm{MSE}(|\hat{{S}}_\ell(k)|^c e^{j\phi(\hat{S}_\ell(k))},\,\, |{S}_\ell(k)|^c e^{j\phi({S}_\ell(k))}), \label{eq:ccomp}
\end{align}
which are then averaged over a training sequence following

\vskip-3pt

\begin{equation}
    J^{\mathrm{xC}} =  10\cdot\log\bigg( {\varepsilon} +  \frac{1}{L^\mathrm{train}}\!\!\!\!\!\!\!\!\!\!\sum_{\ell\in \{1,...,L^\mathrm{train}\}}\!\!\!\!\!\!\!\!\!\! {J}_{\ell}^{\mathrm{xC}}\bigg), \label{eq:log}
\end{equation}
where $J_\ell^{\mathrm{xC}}$ represents either (\ref{eq:magcomp}) or (\ref{eq:ccomp}). We then combine both ${J}^{\mathrm{mC}}$ and ${J}^{\mathrm{cC}}$ to the final loss
({with $\beta=0.7$, chosen based on the optimal performance for the trained models achieved on $\mathcal{D}_\mathrm{syn}^\mathrm{test,MS}$})

\vskip-3pt

\begin{equation}
    J^\mathrm{mcC} = (1-\beta)J^\mathrm{mC}+\beta J^\mathrm{cC}.
\end{equation}

\subsection{Training: Fullband BWE}
\label{subsec:trainBWE}

We perform BWE training as an independent task without any of the previous processing stages present, which allows us to execute this step independently of all previously mentioned training stages. Since $\mathcal{D}_\mathrm{syn}$ does not provide consistent FB speech material, we perform BWE training on the CSTR-VCTK dataset \cite{VCTK}. This dataset was recorded at $96$\,kHz before being downsampled to $48$\,kHz and features consistent high frequency components. A number of 10,000 files of 10\,s length on average are generated by concatenating 3 random speech utterances from the same speaker. The model was trained on this data using a custom amplitude spectrogram loss 
\begin{align}
{J}_\ell^{\mathrm{BWE}} & = 
\mathrm{MSE}({\delta}_\ell(k)\cdot|\hat{S}^\mathrm{UB}_\ell(k)|,\,\,
             {\delta}_\ell(k)\cdot|{S}^\mathrm{UB}_\ell(k)|)\label{eq:bwe}
\end{align}
that penalizes an overestimation in DFT bin $k$ by a factor 
\begin{align}
{\delta}_{\ell}(k) & = \begin{cases}
2 & \text{if } \big|  \hat{S}^\mathrm{UB}_{\ell}(k) \big|  > \big| S^\mathrm{UB}_{\ell}(k) \big|  \\
1 & \, \text{else}.
\end{cases}
\end{align}
Again, frame-based losses (\ref{eq:bwe}) are averaged over time and the logarithm is taken as in (\ref{eq:log}). The training configuration follows the previous steps, except for the initial learning rate being set to $1\!\cdot\!10^{-3}$. Since this DNN is much smaller than the AEC and PF FCRNs, we can afford to use entire files as training sequences, without limiting lengths to $L^\mathrm{train}$.

\section{Results and Discussion}
\label{sec:results}

To evaluate our proposed system, we present three distinct tests. First, we compare our model against the ICASSP 2022 AEC Challenge baseline provided by Microsoft ("Baseline~\cite{Reddy2021}") and against our previous Y$^2$-Net configuration ("Y$^2$-Net FCRN~\cite{seidel21_interspeech}", trained using the same steps as our proposed model) on instrumental {\it wideband} metrics. PESQ~\cite{ITU_P862.2_Corr} is used to determine overall speech quality while the black-box metrics PESQ$_\text{BB}$, ERLE$_\text{BB}$, and dSNR$_\text{BB}$ \cite{fingscheidt_signalseparation, seidel21_interspeech} offer more information about NE speech {\it component} quality and echo suppression. The test set $\mathcal{D}_\mathrm{syn}^\mathrm{test,VCTK}$ contains 1000 files of double-talk scenarios that differ in speakers and echo characteristics from the training data. NE and FE speakers are both taken from the CSTR-VCTK dataset. The FE speaker signal is subject to non-linear distortion \cite{zhang2021} and convolved with a room impulse response generated using the image method~\cite{image_method} to form the echo component at the microphone. Noise is added from the FB dataset of the ICASSP 2022 DNS Challenge~\cite{Reddy2021}. The SER is randomly chosen between -10\,dB and 10\,dB, and the SNR is chosen between 0\,dB and 40\,dB. In Table \ref{tab:synthetic}, we compare the baseline to three different models: our model as described (in bold), a version of it where the AEC training stage was conducted with $L^\mathrm{train}\!=50$ and $B=16$ like all other stages, and a model trained with the same steps, but using the \mbox{Y$^2$-Net} architecture and training targets. Our fully mask-based models outperform both baseline and Y$^2$-Net. While both models show a similar overall performance on PESQ, the $L^\mathrm{train}\!=200$ model preserves NE speech quality (PESQ$_\text{BB}$) better at the {cost of less aggressive echo suppression, resulting in a lower ERLE$_\text{BB}$ score.}

\setlength{\abovecaptionskip}{2mm}

\begin{table}[t]
	\caption{{Instrumental \textit{wideband} ratings} on the {preliminary (VCTK) synthetic testset} $\mathcal{D}_\mathrm{syn}^\mathrm{test,VCTK}$. Best results are marked in {bold}. Our proposed model is marked in {bold}. {Values (50), (200) denote the sequence length (in frames) during AEC stage training.}}
	\label{tab:synthetic}
	\centering
	\setlength\tabcolsep{3.5pt}
	\renewcommand{\arraystretch}{1.00}
	\begin{tabular}{ p{29.9mm}p{0mm}p{8.0mm}<{\centering}p{8.0mm}<{\centering}p{8.0mm}<{\centering}p{8.0mm}<{\centering}}
		\toprule
		\multicolumn{1}{c}{Method:   /   {Metric}:} && \multicolumn{1}{c}{{PESQ}} & \multicolumn{1}{c}{{PESQ$_\text{BB}$}}  & \multicolumn{1}{c}{{ERLE$_\text{BB}$}} & \multicolumn{1}{c}{{$\Delta$SNR$_\text{BB}$}}\\
		\midrule
		Unprocessed 
		&& 2.10 & N/A & N/A & N/A \\
		Baseline~\cite{Reddy2021}
		&& 2.04 & 3.21 & 25.48 & 4.45\\
		{{Y$^2$-Net FCRN (50)}~\cite{seidel21_interspeech}}\! 
		&& 2.14 & 3.19 & 23.93 & 6.58\\
		Ours (50) 
		&& \textbf{2.23} & 3.32 & \textbf{30.40} & \textbf{6.64}\\
		\textbf{Ours (200)} 
		&& 2.20 & \textbf{3.44} & 23.54 & 6.21\\
		\midrule
		{Ours (200) -PF} 
		&& 1.96 & 4.09 & 8.19 & 1.93\\
		\bottomrule
	\end{tabular}
\end{table}

\begin{table}[t]
	\caption{{Instrumental ratings} on the \textit{fullband} {preliminary synthetic testset} $\mathcal{D}_\mathrm{syn}^\mathrm{test,MS}$, partly with omitted processing stages (-PF, -BWE). Asterisk-marked scores refer to signals limited to WB components.}
	\label{tab:synFB}
	\centering
	\setlength\tabcolsep{1pt}
	\renewcommand{\arraystretch}{1.00}
	\begin{tabular}{ p{40mm}p{15mm}<{\centering}p{15mm}<{\centering}}
		\toprule
		\multicolumn{1}{c}{{Method}} & \multicolumn{1}{c}{{POLQA}} & \multicolumn{1}{c}{{ViSQOL}}\\
		\midrule
		Unprocessed & 2.08{/2.02*} & 3.66{/3.33*} \\
		Baseline~\cite{Reddy2021} &  2.35 & 3.59 \\
		Ours (200) -PF -BWE & 2.35 & 3.55 \\
		Ours (200) -BWE & 2.34 & 3.61\\
		\textbf{Ours (200)}  & \textbf{2.36} &  \textbf{3.71}\\
		Ours (50) & 2.29 & 3.65\\
		\bottomrule
	\end{tabular}
\end{table}

The second test is conducted on the test set $\mathcal{D}_\mathrm{syn}^\mathrm{test,MS}$, defined in Section \ref{subsec:trainWB}, and illustrates the effects of each processing stage on the overall FB speech quality measured using POLQA~\cite{ITU_P863} {in super-wideband mode ($32$\,kHz) and ViSQOL~\cite{chinen2020visqol} in audio-mode ($48$\,kHz) to evaluate on the highest possible bandwidth}. Table \ref{tab:synFB} shows that our proposed $L^\mathrm{train}=200$ model outperforms both the baseline and the $L^\mathrm{train}=50$ model. An increase in speech quality can be observed after each stage, with a major improvement caused by the BWE stage. {The differences are more significant on the ViSQOL metric since it emphasizes the loss of upper bands more than POLQA, as can be seen in the score difference between unprocessed data before and after applying a low-pass filter to remove all frequency components above 8 kHz (the latter marked by an asterisk)}.

\begin{table}[t]
	\caption{{\textit{Wideband} ratings} according to the instrumental AECMOS metric on the {(blind) real testset} $\mathcal{D}_\mathrm{real}^\mathrm{test}$.}
	\label{tab:real2}
	\centering
	\setlength\tabcolsep{1pt}
	\renewcommand{\arraystretch}{1.00}
	\begin{tabular}{ p{28mm}p{8.5mm}<{\centering}p{8.5mm}<{\centering}p{8.5mm}<{\centering}p{8.5mm}<{\centering}p{8.5mm}<{\centering}}
		\toprule
		& \multicolumn{2}{c}{{single-talk}} & \multicolumn{2}{c}{{double-talk}}&\\\cmidrule(lr){4-5}\cmidrule(lr){2-3}
		\multicolumn{1}{c}{{Method}} & \multicolumn{1}{c}{{FE}} & \multicolumn{1}{c}{{NE}} & \multicolumn{1}{c}{{FE}} & \multicolumn{1}{c}{{NE}} & \multicolumn{1}{c}{{mean}}\\
		\midrule
		{Unprocessed} 
		& 2.93 & 3.94 & 2.94 & \textbf{3.33} & 3.29
		\\
		{Baseline}~\cite{Reddy2021} & 4.48 & 3.90 & 4.50 & 1.86 & 3.69
		\\
		{\textbf{Ours (200)}}  & \textbf{4.59} & \textbf{3.94} & \textbf{4.62} & 2.03 & \textbf{3.80}\\
		\bottomrule
	\end{tabular}
\end{table}

Our final test in Table \ref{tab:real2} reports the performance achieved on the final blind test set $\mathcal{D}_\mathrm{real}^\mathrm{test}$ of the ICASSP 2022 Acoustic Echo Cancellation Challenge according to the instrumental AECMOS metric~\cite{purin2021aecmos}. $\mathcal{D}_\mathrm{real}^\mathrm{test}$ consists of 800 files containing real-world recordings. Four scenarios are evaluated by predicting scores as they would be given in a subjective listening test: FE single-talk "echo DMOS" test (P.831~\cite{ITU-P831}), NE single-talk "MOS" test (P.808), double-talk "echo DMOS" test (P.831), and double-talk "other degradation DMOS" test (P.831). For double-talk, the "echo DMOS" test will be labeled as "FE", describing the echo annoyance. The "other degradation DMOS" test is dubbed as "NE", describing quality of NE speech. More details can be found in \cite{Sridhar2021}. Please note that these estimations are only computed on the WB signal components, as AECMOS does not offer a FB mode. Our proposed model is able to achieve high scores on the FE metrics while maintaining NE single-talk quality without any degradation. In DT a degradation of the NE signal is observed, but to a significantly lower extent as for the baseline. Overall, our proposed model shows a mean improvement of 0.1 MOS points over the baseline with the further advantage of offering a bandwidth-scalable and modular approach.

\section{Conclusions}
\label{sec:conclusions}

In this work we present a modular and bandwidth-scalable model for acoustic echo cancellation and noise suppression, supporting wideband up to fullband speech with a \textit{single} DNN architecture {for all sampling rates which only requires adaptation of the applied DFT size and swapping the weight set.} The novel fully mask-based approach not only shows an improvement over our previously published Y$^2$-Net architecture, but also outperforms the ICAASP 2022 Acoustic Echo Cancellation Challenge baseline by an average 0.1 MOS points on the respective challenge blind test set.

\bibliographystyle{IEEEbib}
\bibliography{IEEEabrv,MGCabrv,strings,refs}

\end{document}